# AISMOTIF: An Artificial Immune System for DNA Motif Discovery

**Seeja.K.R**

**Department of Computer Science, Jamia Hamdard University,
New Delhi-110062, India**

## Abstract

Discovery of transcription factor binding sites is a much explored and still exploring area of research in functional genomics. Many computational tools have been developed for finding motifs and each of them has their own advantages as well as disadvantages. Most of these algorithms need prior knowledge about the data to construct background models. However there is not a single technique that can be considered as best for finding regulatory motifs. This paper proposes an artificial immune system based algorithm for finding the transcription factor binding sites or motifs and two new weighted scores for motif evaluation. The algorithm is enumerative, but sufficient pruning of the pattern search space has been incorporated using immune system concepts. The performance of AISMOTIF has been evaluated by comparing it with eight state of art composite motif discovery algorithms and found that AISMOTIF predicts known motifs as well as new motifs from the benchmark dataset without any prior knowledge about the data.

**Keywords:** *Artificial immune system, motif discovery, transcription factor binding sites.*

## 1. Introduction

Regulatory motifs play an important role in the expression of genes, as they control the expression or regulation of a group of genes involved in a similar cellular function. Regulatory motifs are short patterns of nucleotides, usually 5-20 bp long, found common in the promoter region of set of co-expressed genes. Identification of these motifs gives insight into the regulatory mechanism of genes. Motif discovery algorithms aim to discover these common patterns, which may present in either strand of the DNA double helix. Identification of DNA motifs is complex due to mutations, which make them weekly conserved patterns.

Biological experiments to find transcription factor binding sites, like ChIP-chip [1] and DNA footprinting [2], are very expensive procedures. Many cheaper computational methods have been proposed as aids to these biological experiments. Survey of motif discovery algorithms [3, 4, 5, 6], classify them broadly into enumerative and probabilistic algorithms. These methods try to discover either consensus string or position weight matrix or position frequency matrix corresponds to the identified motifs. Oligoanalysis[7] is a simple word based algorithm that has been proved efficient for extracting short motifs from yeast regulatory families. Yeast Motif Finder (YMF)[8] is another word based algorithm, which uses z-score and a background model based on Markov chain, to produce the motifs with greatest *z*-scores. There have been algorithms using efficient data structures like suffix trees [9], tables to store intermediate results[10] and using graph theory concepts[11]. Consensus [12] is a greedy probabilistic algorithm that tries to find highest information content motif. The method is based on a position frequency matrix representation of binding site patterns. Most of the probabilistic algorithms use statistical techniques like expectation maximization[13] and Gibbs sampling[14].

Some algorithms based on phylogenetic foot printing[15,16,17] instead of promoter sequences of co-expressed genes have been proposed. The coregulated genes approach requires a reliable method for identifying coregulated genes. In phylogenetic foot printing approach, it is possible to identify motifs specific to even a single gene. The standard method used for phylogenetic footprinting is to construct a global multiple alignment of the orthologous promoter sequences and then identify conserved region in the alignment using a tool such as CLUSTALW[18]. Many hybrid approaches[19,20,21,22] of phylogenetic foot printing and promoter sequences of coregulated genes were also suggested.

Many researchers have explored the use of computational intelligence like evolutionary algorithms [23], genetic algorithm[24] and neural networks[25] in motif discovery. Artificial Immune System (AIS) is a biologically inspired information processing system. It is developed using the theories and components of the natural immune system, which is a highly parallel and distributed adaptive system. According to surveys of artificial immune system [26,27], AIS techniques are mainly based on three theories of immunology-clonal selection [28], immune network[29] and negative selection[30]. CLONALG[31] is a clonal selection algorithm to perform pattern recognition tasks.

Negative selection algorithms[32] were also proposed for applications like anomaly detection and network intrusion detection. Some of the reported applications of AIS are on computer and network security [33,34], recognition of promoter sequences [35] and motif recognition from time series data [36].

This paper describes an enumerative algorithm for regulatory motif discovery based on immune theory. Generally, enumerative algorithms generate all $4^l$ possible combinations of {A, C, G, T} of length $l$. By considering each of this generated sequence as original motif, these algorithms try to find all motifs with a maximum of d mutations of original motif. They are very expensive because of this brute force approach. Since motifs are subsequences present in the input promoter sequences, it is enough to consider the subsequences of length $l$ from the input sequences instead of all $4^l$ combinations. This principle is incorporated in the proposed algorithm using immune system concepts. This algorithm is based on the immune system's ability to distinguish between the foreign cells (nonself) and the cells of the body (self).

## 2. Methods

### 2.1. Artificial Immune System

The natural immune system defends the body against harmful diseases and infections. It is capable of recognizing any foreign cell (pathogen) and eliminating it from the body. The immune system performs a kind of pattern recognition to distinguish the foreign cells and the cells of the body. Natural immune system has the ability of learning about foreign substances that enter the body and produces the antibodies, which can attack the antigens associated with the pathogens. The immune system protects body by learning and memory. If the immune system detects a pathogen that it has not encountered before, it undergoes a primary response, during which it learns the structure of the specific pathogen and create an antibody for that pathogen. The immune system maintains a *memory* of infections so that if a pathogen already attacked is encountered, it can respond quickly by activating the already produced antibody which is stored in the memory. Thus the secondary immune response occurs, when the same antigen is encountered again. After successful recognition, the adaptive immune response is elicited. The immune system reproduces those cells capable of recognizing and binding with antigens. The cellular reproduction in the immune system is based on cloning. Cloning is the creation of offspring cells that are copies of their parent cells subject to mutations. Due to the mutations, the cells within a clone are all similar but present slight differences and are capable of recognizing the antigen that triggered the immune response. A selective mechanism guarantees that those offspring cells that better recognize the antigen (affinity maturation), which elicited the response, are selected for cloning. These cells are called memory cells. Affinity refers to the degree of binding of the cell receptor with the antigen. The higher the affinity the stronger the binding and thus the better the immune recognition and response. The whole process of antigen recognition, cell proliferation and differentiation into memory cells is named as clonal selection.

Artificial Immune System (AIS) is a collection of techniques originated from the theory of immunology. The pattern recognition applications are mainly based on the clonal selection principle of immune theory. The pattern matching algorithms use the clonal selection for learning pattern detectors through cloning, mutation and selection phases. They maintain a memory of pattern detectors using some efficient data structures. When a new pattern is to be identified, the memory detector that matches more with the incoming pattern is cloned and mutated to accommodate the new pattern. This becomes the memory detector of the new pattern and it is stored in the immune memory.

### 2.2. AISMOTIF Algorithm

This algorithm is based on immune memory and clonal selection. During the learning phase initial memory detectors are identified and stored in the memory. The immune memory is maintained as a table called *memory detector table*. AISMOTIF finds motifs of given length $l$ from the set of promoter sequences of co-expressed genes. Thus the input of the algorithm is a file containing the promoter sequences and the motif length $l$. The algorithm does not need any background sequence. The AISMOTIF algorithm described in this section assumes that there is at least one occurrence of a motif in every input sequence. This assumption reduces the size of the immune memory considerably. However the algorithm can be easily extended for other conditions like zero or one occurrence per sequence. The initial memory detectors (antibodies) are created by generating all subsequences of length $l$ from a sequence selected randomly form the input set of promoter sequences. These subsequences are created by using sliding window approach. Since at least one motif is present in every sequence according to our assumption all these subsequences can be considered as candidate motifs. Now select one of the sequences from the remaining sequence and generate subsequences of length $l$ by using sliding approach. These subsequences correspond to antigens. For each of these antigens find the antibody which gives the best match or affinity according to the *Match Score* defined in section 2.2.1. Make as many clones of antibodies as many matches are there. Then mutate each clone by attaching the matched subsequence and remove the antibodies that are not matching with any of the antigens from the memory. Now all memory detectors in the table have two subsequences of length *l*. Again select one of the sequences from the remaining sequences and repeat the whole process for each of the input sequences. At the end, the memory detector table has a set of entries each representing a motif. If there are n input sequences then each entry in the memory detector table has n subsequences one from each of the input sequences. In order to find the best motifs, sort the memory detector table based on the *Information Score* of the motifs. A new *Information Score* is proposed in section 2.2.2.

### 2.2.1 Match Score

The proposed *Match Score* is a weighed one. If the presence of a nucleotide in the subsequence to be matched with the already identified motif instances in the memory detector table increases the count of the top nucleotide in that position, then a high weight is given to that position. To assign weights, the count value is multiplied with weight, which is equal to count/total number.

Let C is the position count matrix of the motif instances in memory detector table and M is the subsequence to be matched

$$MatchScore = \frac{1}{(n+1)^2 l} \sum_{i=1}^{l} \left( C_{i, M_i + 1} \right)^2$$

where $M_i$ is the nucleotide at position $i$ in the subsequence, $n$ is the number of motif instances in memory detector and $l$ is the length of the motif. The match score is a value between 0 and 1, where 1 represents exact match.

For example, let GATCACCG, GATTACCG, GATTACCG are the motif instances in the memory detector and GATTAACG is the subsequence.

Position Count Matrix, C

|   | A | T | G | C |
|---|---|---|---|---|
| 1 | 0 | 0 | 3 | 0 |
| 2 | 3 | 0 | 0 | 0 |
| 3 | 0 | 3 | 0 | 0 |
| 4 | 0 | 2 | 0 | 1 |
| 5 | 3 | 0 | 0 | 0 |
| 6 | 0 | 0 | 0 | 3 |
| 7 | 0 | 0 | 0 | 3 |
| 8 | 0 | 0 | 3 | 0 |

M= GATTAACG, *l*=8, *n*=3

MatchScore=$(1/128)*(4^2+4^2+4^2+3^2+4^2+1^2+4^2+4^2)$
            =106/128=**0.828**

### 2.2.2 Information Score

The proposed information content score is based on the similarity of nucleotides at each position of the motif. The more the motif instances are similar the more its information content. Here also a weight is assigned as in case of match score. Highest $C_{i,j}$ value in each row is added to get the information content. The weight value is highest $C_{i,j}$ divided by number of input sequences.

$$InformationScore = \frac{1}{n^2 l} \sum_{i=1}^{l} \left( \max_{j \in N} C_{i,j} \right)^2$$

where N={A,T,G,C} and n is the number of input sequences. The information score is a value between 0 and 1(weakest motif to strongest motif).

For example, consider the same motif instances and position count matrix in section 2.2.1.

Information Score = $(1/72)*(3^2+3^2+3^2+2^2+3^2+3^2+3^2+3^2)$
            = 67/72=**0.93**

### 2.2.3 Pseudo code

*Input: Set of promoter sequences and length of motif l*

*Output: Set of motif instances*

*AISMOTIF(L,l )*

```
{
 Randomly select any one promoter sequence;
 Create antibodies(memory detectors) by generating all
        subsequence of length l using  sliding window
        approach;
 Create immune memory by storing these antibodies
        in table MT;
 for each of the remaining n-1 sequences do
  {
   select the next sequence;
   create antigens by generating subsequences of length l;
   for each antigen s do{
        find the antibody which best matches the antigen by
            finding  MatchScore(MT[i],s);
        Clone the antibody and mutate it by appending the
            antigen sequence with the antibody sequence;
        This becomes the new antibody for the new antigen
            and  stores it in immune memory;
   }
   Delete those antibodies from immune memory which did
            not match with any of the antigens;
  }
}
```

### 2.2.4 Multiple motifs

The pseudo code of the AISMOTIF given in section 2.2.3 assumes one occurrence per sequence (oops). Therefore it considers multiple occurrence of a motif in the same sequence as different motif instances and this result in the identification of same motif many times. A simple modification of the basic AISMOTIF can dealt with this problem. Add a subsequence to the associated list of an already stored subsequence if its match score is greater than or equal to the user defined minimum match threshold, instead of storing it as a separate entry in immune memory MT.

### 2.2.5 Identification of motifs from both strands of DNA

A good motif discovery algorithm has to identify motifs which may present in either strand of the DNA double helix. This can be done by modifying the create antibody and create antigen step of AISMOTIF algorithm by storing the reverse complement of the subsequences along with the subsequence.

### 2.2.6 Variable length motifs

Motif length is one of the inputs of AISMOTIF. In order to get motifs of different lengths, modify the algorithm by adding an outer for loop whose control variable changes from minimum length to maximum length.

### 2.2.7 Search for known motifs

The algorithm can be modified for finding known motifs (clinically identified motifs), by creating the initial antibody set with the known motifs.

## 3. Results

The AISMOTIF algorithm has been implemented in Perl and its performance is evaluated with benchmark dataset[6] available at Norwegian University of Science and Technology website (http://tare.medisin.ntnu.no/). Since AISMOTIF identifies many motifs simultaneously, the composite motif benchmark assessment tool provided in the website is selected to compare the performance of AISMOTIF with the existing composite motif discovery algorithms. In order to evaluate the performance of AISMOTIF, all the three datasets (TRANSCompel, Liver and Muscle) are applied to the AISMOTIF algorithm and the generated outputs are uploaded to the database of the benchmark assessment tool. The web service provided various assessment statistics like nucleotide correlation coefficient (nCC), sensitivity (Sn), specificity (Sp), average site performance (ASP), performance coefficient (PC) and positive predictive value (PPV). The performance of the AISMOTIF is compared with the performance of eight composite motif discovery tools namely CMA[37], ModuleSearcher[38], Stubb[39], MSCAN[40], MCAST[41], Cister[42], Cluster-Buster[43] and CisModule[44].

### 3.1. TRANSCompel dataset

It is a collection of 10 datasets - AP1-Ets, AP1-NFAT, AP1-NFkappaB, CEBP-NFkappaB, Ebox-Ets, Ets-AML, IRF-NFkappaB, NFkappaB-HMGIY, PU1-IRF and Sp1-Ets. AISMOTIF could identify many occurrences of the motifs known to be present in these datasets under the assumption of one occurrence per sequence. The number of motifs identified by AISMOTIF from these datasets is 175, 168, 244, 187, 334, 270, 254, 224, 368 and 221 respectively. The performance comparison of AISMOTIF with other motif discovery tools on TRANSCompel dataset is given in table 1.

Table 1: Performance Comparison of AISMOTIF with other motif discovery tools on TRANSCompel dataset

| Methods | ASP | nCC | SP | PC | Sn | PPV |
|---|---|---|---|---|---|---|
| AISMO TIF | **0.499** | 0.035 | 0.099 | 0.042 | **0.956** | 0.042 |
| CMA | **0.487** | 0.327 | 0.849 | 0.194 | **0.772** | 0.201 |
| ModuleSearcher | **0.394** | 0.346 | 0.98 | 0.235 | **0.343** | 0.445 |
| Stubb | **0.412** | 0.206 | 0.752 | 0.108 | **0.712** | 0.113 |
| MSCAN | **0.347** | 0.299 | 0.994 | 0.2 | **0.243** | 0.45 |
| MCAST | **0.46** | 0.262 | 0.785 | 0.138 | **0.778** | 0.143 |
| Cister | **0.301** | 0.217 | 0.961 | 0.124 | **0.292** | 0.311 |
| Cluster-Buster | **0.412** | 0.332 | 0.937 | 0.223 | **0.524** | 0.301 |
| CisModule | **0.139** | 0.008 | 0.777 | 0.037 | **0.233** | 0.044 |

## 3.2. Muscle dataset

This dataset contain 24 sequences of which 10 sequences were from the mouse genome, 6 from human, 5 from rat, 2 from chicken and 1 from cow. Five motifs (Mef-2, Myf, Sp1, SRF and Tef) were reported as important in muscle regulation. AISMOTIF could identify many occurrences of Sp1, SRF, Mef-2, Myf and Tef along with other motifs under the assumption of one occurrence per sequence. The AISMOTIF generated 97 motif instances simultaneously, sorted in the descending order of information score. The performance comparison of AISMOTIF with other motif discovery tools on Muscle dataset is given in table 2.

Table 2: Performance Comparison of AISMOTIF with other motif discovery tools on Muscle dataset

| Methods | ASP | nCC | SP | PC | Sn | PPV |
|---|---|---|---|---|---|---|
| AISMOTIF | **0.454** | 0.098 | 0.383 | 0.143 | **0.759** | 0.149 |
| CMA | **0.533** | 0.462 | 0.922 | 0.362 | **0.559** | 0.507 |
| CisModule | **0.488** | 0.289 | 0.695 | 0.231 | **0.724** | 0.253 |
| ModuleSearcher | **0.526** | 0.463 | 0.948 | 0.354 | **0.483** | 0.57 |
| Stubb | **0.443** | 0.243 | 0.699 | 0.209 | **0.65** | 0.236 |
| MSCAN | **0.57** | 0.498 | 0.914 | 0.393 | **0.629** | 0.512 |
| MCAST | **0.584** | 0.296 | 0.484 | 0.208 | **0.958** | 0.21 |
| Cister | **0.588** | 0.356 | 0.613 | 0.249 | **0.923** | 0.254 |
| Cluster-Buster | **0.547** | 0.411 | 0.804 | 0.313 | **0.743** | 0.352 |

## 3.3. Liver dataset

This dataset contained 12 sequences of which 8 sequences were from human, 2 from rat and the last 2 from mouse and chicken. Four motifs (C/EBP, HNF-1, HNF-3 and HNF-4) were reported as important in liver regulation. AISMOTIF could identify many occurrences of all these motifs along with other motifs under the assumption of one occurrence per sequence. The AISMOTIF generated 203 motif instances simultaneously, sorted in the descending order of information content score. The performance comparison of AISMOTIF with other motif discovery tools on Liver dataset is given in table 3.

Table 3: Performance Comparison of AISMOTIF with other motif discovery tools on Liver dataset

| Methods | ASP | nCC | SP | PC | Sn | PPV |
|---|---|---|---|---|---|---|
| AISMOTIF | **0.506** | 0.014 | 0.116 | 0.113 | **0.898** | 0.114 |
| CMA | **0.461** | 0.358 | 0.867 | 0.278 | **0.571** | 0.352 |
| CisModule | **0.142** | -0.005 | 0.819 | 0.072 | **0.175** | 0.109 |
| ModuleSearcher | **0.493** | 0.425 | 0.975 | 0.291 | **0.348** | 0.638 |
| Stubb | **0.548** | 0.476 | 0.913 | 0.368 | **0.621** | 0.475 |
| MSCAN | **0.591** | 0.51 | 0.99 | 0.332 | **0.357** | 0.824 |
| MCAST | **0.615** | 0.504 | 0.846 | 0.372 | **0.826** | 0.404 |
| Cister | **0.448** | 0.306 | 0.802 | 0.24 | **0.614** | 0.282 |
| Cluster-Buster | **0.638** | 0.588 | 0.944 | 0.466 | **0.673** | 0.602 |

It is found that the *sensitivity score* of AISMOTIF is high compared to other motif discovery tools tested here. The sensitivity score gives the fraction of known sites that are predicted. At the same time AISMOTIF makes a lot of false positive predictions indicated by the lower *positive predictive values*. The positive predictive value gives the fraction of the predicted sites that are known. AISMOTIF gives many motifs which are unknown. For example, the liver dataset contains only 4 known sites. But the AISMOTIF predicted 203 motif instances (some of them are duplicate motifs as described in section 2.2.4). These unknown motifs generated by AISMOTIF may be unannotated true positives; even though the benchmark assessment tool considers them as false positives. The authors of the benchmark assessment tool (Klepper et al 2008) clearly mentioned this fact. The lower values of the *performance coefficient*, *specificity* and hence that of the *correlation coefficient* is due to these false positives. The *average site performance* depends on sensitivity score and hence its value is also high.

The size of the datasets and the time taken by AISMOTIF (on a PC with T7200 processor with 2GHz speed and 1 GB RAM) for motif length =12 are given in table 4. The longest sequence in each of the dataset is 1000bp. It is found that the time taken by the AISMOTIF to generate the top motifs is quite reasonable compared to other enumerative algorithms due to the immune system based pruning. The running time of AISMOTIF could not be compared with other motif discovery tools because none of the tools are generating these many motifs simultaneously.

Table 4: Running time of AISMOTIF on different benchmark datasets

| Data Set | No. of sequences | Smallest sequence (bp) | Total Size (bp) | **Running time(sec)** |
|---|---|---|---|---|
| AP1-Ets | 16 | 609 | 14860 | **2318** |
| AP1-NFAT | 8 | 294 | 6893 | **1031** |
| AP1-NFkappaB | 8 | 532 | 6532 | **1392** |
| CEBP-NFkappaB | 8 | 308 | 7308 | **1147** |
| Ebox-Ets | 4 | 492 | 3489 | **672** |
| Ets-AML | 5 | 480 | 4053 | **505** |
| IRF-NFkappaB | 6 | 596 | 5344 | **964** |
| NFkappaB-HMGIY | 6 | 393 | 5394 | **875** |
| PU1-IRF | 5 | 722 | 4530 | **1058** |
| Sp1-Ets | 7 | 309 | 5787 | **1000** |
| Liver | 12 | 943 | 11943 | **2132** |
| Muscle | 24 | 269 | 20427 | **2255** |

## 4. Conclusion

In this paper, an artificial immune system based pattern discovery algorithm for finding regulatory motifs from promoter sequences of a set of co-expressed genes have been proposed and demonstrated. The major advantage of this algorithm is its ability to generate all possible motifs in the input sequences simultaneously. It does not need any complicated parameter settings like other existing algorithms since the only inputs to this algorithm are the input promoter sequences and the length of motif. It also does not need any kind of background model and any knowledge of background distribution of nucleotides and the species. A new weighted match score and an information score independent of any background probability have also been proposed. Since there is no chance of local optimum and the proposed AIS based enumeration is not so time consuming, AISMOTIF is a good choice for *de novo* motif discovery.

## References

[1] Horak CE, Snyder M (2002) ChIP-chip: a genomic approach for identifying transcription factor binding sites *Methods Enzymol.*, 350, 469-83

[2] Gunderson SI, Murphy JT, Knuth MW, Steinberg TH, Dahlberg JH, Burgess RR. (1988) Binding of transcription factors to the promoter of the human U1 RNA gene studied by footprinting. *J Biol Chem.*, 263(33), 17603-10.

[3] Hu, Jianjun, Li, Bin Kihara, Daisuke (2005) Limitations and potentials of current motif discovery algorithms. *Nucl. Acids Res.* ,33(15), 4899-4913.

[4] M. Tompa, N. Li, T. L. Bailey , G. M. Church , B. De Moor, E. Eskin, A. V. Favorov, M. C. Frith, Y. Fu, W. J. Kent, V. J. Makeev, A. A. Mironov, W. S. Noble, G. Pavesi, G. Pesole, M. Regnier, N. Simonis, S. Sinha, G. Thijs, J. van Helden, M. Vandenbogaert, Z. Weng, C. Workman, C. Ye, and Z. Zhu (2005) Assessing Computational Tools for the Discovery of Transcription Factor Binding Sites. *Nature Biotechnology*, 23(1), 137 – 144.

[5] Geir Kjetil Sandve, Finn Drabløs (2006) A survey of motif discovery methods in an integrated framework. *Biology Direct* , 1:11.

[6] Klepper K, Sandve GK, Abul O,Johansen J, Drablos F (2008) Assessment of composite motif discovery methods *BMC Bioinformatics,* 9():123

[7] Helden, J. V., Andre, B, Collado-Vides, J.(1998) Extracting regulatory sites from the upstream region of yeast genes by computational analysis of oligonucleotide frequencies *J Mol Biol*, 281 (5), 827-842.

[8] Sinha, S., Tompa, M.(2002) Discovery of novel transcription factor binding sites by statistical overrepresentation. *Nucleic Acids Res*, 30(24), 5549-60.

[9] Sagot, M.F(1998) Spelling approximate repeated or common motifs using a sufix tree. *In Proceedings of the Third Latin American Symposium on Theoretical Informatics* ,374-390.

[10] Seeja,K,R, Alam,M,A, Jain,S,K (2009) MotifMiner: A Table Driven Greedy Algorithm for DNA Motif Mining. *Emerging Computing Technology and Applications. with Aspects of Artificial Intelligence*, LNAI 5755, 397–406.

[11] Pevzner P.A.,Sze S,Combinatorial Approaches to finding subtle signals in DNA sequences. Proceedings of the International Conference on Intelligent Systems for Molecular Biology. AAAI Press; 2000. p. 269-278

[12] Hertz GZ, Hartzell GW, Stormo GD(1990) Identification of consensus patterns in unaligned DNA sequences known to be functionally related. *Comput Appl Biosci* 6, 81-92.

[13] Bailey TL, Elkan C. (1995) Unsupervised learning of multiple motifs in biopolymers using expectation maximization. *Machine Learning,* 21, 51-80.

[14] Liu X, Brutlag DL, Liu JS(2001) BioProspector: discovering conserved DNA motifs in upstream regulatory regions of co-expressed genes. *Proceedings of the Sixth Pacific Symposium on Biocomputing* , 127-138.

[15] McCue L, Thompson W, Carmack C, Ryan M, Liu J, Derbyshire V, Lawrence C (2001) Phylogenetic footprinting of transcription factor binding sites in proteobacterial genomes. *Nucleic Acids Res.*, 29,774-782.

[16] Berezikov, E., Guryev,V., Plasterk, RHA., Cuppen, E. (2004) CONREAL:Conserved regulatory elements anchored alignment algorithm for identification of transcription factor binding sites by phylogenetic footprinting *Genome Research* , 14, 170-178.

[17] Cliften, P, Sudarsanam, P, Desikan, A, Fulton, L, Fulton, B, Majors, J, Waterston, R, Cohen, BA, Johnston, M (2003) Finding functional features in Saccharomyces genomes by phylogenetic footprinting. *Science*, 301, 71-76.

[18] Thompson JD, Higgins DG, Gibson TJ (1994) CLUSTALW: improving the sensitivity of progressive multiple sequence alignment through sequence weighting, position-specific gap penalties and weight matrix choice. ,*Nucleic Acids Res.*, 22(22), 4673-80.

[19] Wang T, Stormo GD (2003) Combining phylogenetic data with coregulated genes to identify regulatory motifs *Bioinformatics* , 19, 2369-2380.

[20] Sinha S, Blanchette M, Tompa M (2004), PhyME: A probabilistic algorithm for finding motifs in sets of orthologous sequences. *BMC Bioinformatics*, 5, 170.

[21] Moses A, Chiang D, Eisen M(2004) Phylogenetic motif detection by expectation-maximization on evolutionary mixtures. *Proceedings of the Ninth Pacific Symposium on Biocomputing* , 324-335.

[22] Siddharthan R, Siggia ED, van Nimwegen E (2005) PhyloGibbs: A Gibbs sampling motif finder that incorporates phylogeny. *PLoS Comput Biol*,, 1, 534-556.

[23] Michael A Lones and Andy M. Tyrrell(2007) Regulatory Motif Discovery Using a Population Clustering Evolutionary Algorithm. *IEEE/ACM Transactions on Computational Biology and Bioinformatics*, 4(3) , 403-414.

[24] Liu FFM, Tsai JJP, Chen RM, Chen SN, Shih SH(2004) FMGA: finding motifs by genetic algorithm. *Fourth IEEE Symposium on Bioinformatics and Bioengineering* , 459-466.

[25] Liu D, Xiong X, Das Gupta B, Zhang H(2006) Motif discoveries in unaligned molecular sequences using self-organizing neural network. *IEEE Transactions on Neural Networks*, 17, 919-928.

[26] Dasgupta, D, Attoh-Okine,N (1997) Immunity-Based Systems: A Survey. *IEEE International Conference on Systems, Man and Cybernetics*, 1, 363-374.

[27] Timmis J., Knight T., De Castro L.N.,Hart E(2004) An overview of artificial immune systems. *Computation in Cells and Tissues: Perspectives and Tools for Thought, Natural Computation Series,*51-86.

[28] Ada, G.L. and Nossal, G.J.V. (1987) The clonal selection theory. *Scientific America,*. 257(2),62-69.

[29] Jerne,N,K (1974) Towards a Network Theory of the Immune System., *Ann. Immunol. (Inst. Pasteur)*, 125C, 373-389.

[30] Nossal GJ.(1994) Negative selection of lymphocytes. *Cell.* 76(2) , 229–239

[31] De Castro, L. N.,Von Zuben, F. J. (2000) The Clonal Selection Algorithm with

[32] Engineering Applications. *GECCO'00 – Workshop Proceedings*, 36-37.

[33] Forrest, S., A. Perelson, Allen, L., Cherukuri, R. (1994) Self-Nonself Discrimination in a Computer. *Proc. of the IEEE Symposium on Research in Security and Privacy*, 202-212.

[34] Oda,T, White, T (2003) Developing an immunity to spam. *Proceedings of the Genetic and Evolutionary Computation Conference*, 231-242.

[35] Schaust, S. and Szczerbicka, H. (2008) Artificial immune systems in the context of misbehavior detection. *Cybern. Syst.,* 39, 136-154.

[36] Cooke,DE, Hunt,JE (1995) Recognising promoter sequences using an artificial immune system. *Proc Int Conf Intell Syst Mol Biol.*, 3, 89-97.

[37] William Wilson, Phil Birkin and Uwe Aickelin (2007) Motif Detection Inspired by Immune Memory *Artificial Immune Systems*,4628/200, 276-287.

[38] Kel AE, Konovalova T, Waleev T, Cheremushkin E, Kel-Margoulis OV, Wingender E(2006) Composite Module Analyst: a fitness-based tool for identification of transcription factor binding site combinations. *Bioinformatics*, 22(10),1190-1197.

[39] Aerts, S, Van Loo, P, Thijs, G, Moreau, Y, De Moor, B (2003) Computational detection of cis-regulatory modules. *Bioinformatics*,19(suppl 2), ii5-14.

[40] Sinha S, van Nimwegen E, Siggia ED (2003) A probabilistic method to detect regulatory modules *Bioinformatics*, 19(Suppl.1), i292-i301.

[41] Johansson, Alkema WBL, Wasserman WW, Lagergren J (2003) Identification of functional clusters of transcription factor binding motifs in genome sequences: the MSCAN algorithm *Bioinformatics*, 19(suppl 1), i169-i176.

[42] Bailey TL, Noble WS (2003) Searching for statistically significant regulatory modules. *Bioinformatics*, 19(Suppl. 2),ii16-ii25.

[43] Frith MC, Hansen U, Weng Z (2001) Detection of cis-element clusters in higher eukaryotic DNA. *Bioinformatics*, 17(10), 878-889.

[44] Frith MC, Li MC, Weng Z (2003) Cluster-buster: finding dense clusters of motifs in DNA sequences" *Nucl Acids Res*, 31(13), 3666-3668.

[45] Zhou Q, Wong WH (2004) CisModule: de novo discovery of cis-regulatory modules by hierarchical mixture modeling. *Proc Natl Acad Sci USA* , 101(33), 12114-12119.

**Seeja.K.R.** received her B.E(Computer Engineering),M.E(Computer Science & Engineering) and PhD (computer Science) degrees in 1995, 2001 and 2010 respectively. Currently, she is assistant professor in the Department of Computer Science of Jamia Hamdard University , New Delhi, India. Her research interests are data mining, Algorithm Design and Artificial Intelligence. She has published 13 research papers in referred journals and conference proceedings.